# Cause of $\frac{1}{f}$ Noise: The Surge of Background Energy

Sheng-Long Xu

Kunming Institute of Physics, Yunnan, P. R. China, 650223

**Abstract:** Based on analyzing various physical examples for 1/f noise we found that the 1/f noise is caused by generating conditions as 'resonance' in acoustics and the effect of matter carrier is secondary. All the physical reasons are summarized into "The surge of background energy", then the 1/f noise is separated from matter carrier. Therefore the attention is focus on the general characters of 1/f noise and the following viewpoints and results are of universal significance. The work 'cause for 1/f Noise' came to a universal relationship:

$$s(f) \propto \frac{|\log f|}{f}$$

All the questions about 1/f noise were solved completely and systematical. Follow-up study indicates that the crux of 1/f noise is a special 'phase' whose entropy is about $10^{-20}$ erg/k.

## 1. What is the *1/f* noise？

What is the *1/f* noise？Japanese scientist 高安秀树[11] indicated "What is called *1/f* noise is the general name of an oscillation whose power spectrum is in inverse proportion to the frequency *f* and also called flicker noise or pink noise. People have found the *1/f* noise in variety phenomenon and worked hard to explore the origin but still not to the end.

As well know the power spectrum of white noise in electric circle is in proportion to $f^0$ and called thermal noise. *1/f* noise would be observed in the circle consisted of either tubes or transistors while *f* is small. Power spectrum in proportion to the frequency $f^{-1}$ means that the longer wave-length of oscillation, the larger amplitude and the oscillation is affect -ed by its history long time ago. It seems that all of these as above were un-accepted and it were imagined that the *1/f* noise would disappear with too small *f*. But the power spectrum of *1/f* noise does not change no matter how long the observation time in experiment. Therefore the power spectrum of *1/f* noise is trust even *f* goes to 0, which just like the relaxation process with a long tail.

In addition to electrical circle, *1/f* noise also observed in various phenomenon such as crystal wafer oscillation, car flow on the high way, variance of temperature with seasons, music, pulses of people, membrane voltage of nerve fiber *et al*. Though it is assume there is a general origin mechanism about *1/f* noise, but still have not satisfied explanation up to now. …"

*1/f* noise is neither mechanical wave nor electromagnetic wave and other knew oscillation but a new un-known origin and universal dynamic form.

The tricky *1/f* noise makes people puzzled for a long time. All the power spectra of $\frac{1}{f}$ noise show the same form [1][2][3][4][5][6] in the low frequency region

$$S(f) \propto \frac{1}{f^M} \qquad (1)$$

The value of parameter $M$ runs in a range which was different for variant studies:

$$0.5 < M < 1.5 \quad [3]$$
$$0.8 < M < 1.2 \quad [7] \qquad (2)$$
$$0.9 < M < 1.4 \quad [8]$$

How understand the power spectra of $\frac{1}{f}$ noise and the value of parameter $M$ as above? Quite a few outstanding scientists have explored these questions. There exist still a lot of questions as follows:

(1) $\frac{1}{f}$ noise can be observed in variety systems no matter what thy are, such as macro or micro, nature or society, simple or complex *et al*. Why?

(2) All the $\frac{1}{f}$ noise follow the same law (1), why? There must be a common origin. What is it?

(3) The values of parameter $M$ is special and difference from the powers of power series in other physics equation. What the reason is?

(4) In fact, what is called constant $M$ is not fixed and separated values, but runs in a continuous region where is open or half open.

(5) The region existed $M$ will be difference in different measurements, which one is reliable?

(6) Among the various values of $M$, which one is true?

(7) The frequency $f$ varies continuously in equation (2), whether it can be broken?

...

Strictly speaking, question (1) and (2) are essential, the others technical. They are difficulty to be solved and even contradictory with each other.

Questions are incisive, bright and clear, which make scientists pay huge attention to and suggest a variety of hypothesis. However it was unexpected that things were so abstruse and difficult that no dawn of solution even through great efforts of several generations of intellectuals. Some people turn to specific systems then explore general mechanism and got certain progress, however overall, it was limited. It is the heart of the matter, that the attention is only focus on inside of objects. The applicability of theorem is limited. For example, it may be suitable for explaining the $\frac{1}{f}$ noise of a device but not for music and meteorology. In fact there are several explanations even for the $\frac{1}{f}$ noise of dc SQUID devices. Clark etc. believe that it is the fluctuation of critical current induced by the fluctuation of temperature while Rogers and Buhrman[9] consider that it is induced by the fluctuation of barrier tunnel conductance.

In general, have studied for several decades the results are still not satisfied which means that prevalent way has been in a tight corner. As one review[8] said "From published references, no any theoretical model is successful and perfect up to now." and "Although this noise has been observed in common and studied since 60 years ago, the origin is still not clear."[11].

We have studied $\frac{1}{f}$ noise for decade. In the beginning, as others we focused attention only on inside of objects to find the result, and failed, then turned to a new way. Unexpectedly, all the questions were solved. The $\frac{1}{f}$ noise is either special appeared or extreme hided which depends mainly on the surroundings and matter carrier subordinately. Therefore the matter carrier was separated from system and the attentions were focused mainly on the general characters. Finally, we have to confirm that the origin of $\frac{1}{f}$ noise is the interaction between system and random effect. According to this thought, all the questions will be solved. Moreover we find the abundant connotation of $\frac{1}{f}$ noise. There are infinite forms for $\frac{1}{f}$ noise and only a small part has been dealt with up to now. The $\frac{1}{f}$ noise possesses exact law and a universal equation will be suitable for all forms of $\frac{1}{f}$ noise:

$$S(f) \propto \frac{|\log f|}{f} \tag{3}$$

## 2. Basic theory

Let us suppose a system which interacts with outside and transfers energy and mass but maintains the elementary characters i.e. the characteristic frequency of system. In fact this system is just a mathematics abstraction for device, music and meteorological phenomena etc. For the sake of convenience we divide the system into $N$ units and $m_k$, ($k=1,2,3\cdots N$) and $\vec{u_k}$ stand for the mass and velocity of unit $k$ separately. The statistical weight per unit is equal no one special. $N$ is so large that $\vec{u_k}$ is uniform inside unit and $m_k \propto \dfrac{1}{N}$.

There are a variety of interactions about system with surroundings, but only the random action induces 1/f noise. Random interference is a weak action force and unpredictable. However the random force $\vec{\wp}(t)$ is limited in time, space and energy, namely it is finite. Especially finite energy makes it be able to carry out Fourier transform. For a particular $\vec{\wp}(t)$ we have

$$\vec{\wp}(t) = \int_{-\infty}^{+\infty} \vec{B}(f) e^{i\omega t} df \tag{4}$$

where $\omega = 2\pi f$.

If neglect dimension, $\vec{B}^*(f)\vec{B}(f)$ is just the power spectrum of $\vec{\wp}(t)$. Then our attention will focus on this power spectrum.

$$\vec{B}^*(f)\vec{B}(f) = \dfrac{\vec{L}^*\vec{L}}{\beta^2 + f^{2\alpha} g(f)} \tag{5}$$

$\alpha > 1$ and $g(f) > 0$ are necessary for maintaining definite energy.

Suppose the system undergoes a sine force $\vec{A}e^{i\lambda t}$ which may be ether intentional such as operating voltage or not or both. The $k$th unit bears $\vec{A_k} e^{i\lambda t}$ and $\vec{A_k} \propto \dfrac{1}{N}$. The velocity of the unit of mass $m_k$ changed from $\vec{u_k}$ to $\vec{u_k} + \delta\vec{u_k}$, we have the motion equation:

$$m_k \dfrac{d}{dt}\left(\delta\vec{u_k}\right) = \vec{A_k} e^{i\lambda t} + \int_{-\infty}^{+\infty} \vec{B_k}(f) e^{i\omega t} df \tag{6}$$

then find the solution:

$$\delta\vec{u_k} = \dfrac{\vec{A_k}}{i\lambda m_k} e^{i\lambda t} + \int_{-\infty}^{+\infty} \dfrac{\vec{B_k}(f)}{i\omega m_k} e^{i\omega t} df \quad . \tag{7}$$

Moreover we get the energy increment:

$$\dfrac{1}{2} m_k \left|\delta\vec{u_k}\right|^2 = \dfrac{\left|\vec{A_k}\right|^2}{2m_k \lambda^2} + \dfrac{1}{2m_k}\left|\int_{-\infty}^{+\infty}\dfrac{\vec{B_k}(f)}{\omega} e^{i\omega t} df\right|^2 + \left\{\int_{-\infty}^{+\infty} \dfrac{\vec{A_k}^* \vec{B_k}}{2m_k \lambda \omega} e^{i(\omega-\lambda)t} df + Co.\right\} \tag{8}$$

Summing for all k units we have:

$$W = \sum_k \frac{1}{2} m_k \left|\vec{\delta u_k}\right|^2 = E + P + \left\{ \int_{-\infty}^{+\infty} S(f) e^{i(\omega-\lambda)t} df + Co. \right\} \qquad (9)$$

*Co.* Stand for conjugate term.
Where

$$\begin{cases} E = \sum_k \dfrac{\left|\vec{A_k}\right|^2}{2 m_k \lambda^2} \\ P = \sum_k \dfrac{1}{2 m_k} \left| \int_{-\infty}^{+\infty} \dfrac{\vec{B_k}(f)}{\omega} e^{i\omega t} df \right|^2 \\ S(f) = \dfrac{1}{2\pi f \lambda} \sum_k \dfrac{\vec{A_k}^* \vec{B_k}(f)}{m_k} \end{cases} \qquad (10)$$

It is explicit that *W* is the total increment of system energy caused by interaction, which includes *E* the action energy of force $\vec{A}e^{i\lambda t}$, *P* the background noise energy induced by random action while *S(f)* is the energy spectrum of the other additional energy,. Equation (9) is the equilibrium equation of energy.

Because *N* is so large that the contribution of per unit to energy, namely, per term of sum is much small, to add or subtract a few term in the sum *S(f)* will not affect more on the finial results. But the both side of equal-sign in equation (9) will arise a small difference which means energy flow existing as well as mass flow. Hence the system changes from closed to open and transfers mass as well as energy with outside e.g. music and meteorology systems. "Keep invariable" means that the mass and energy may transfer but their quantities in the system maintain constant. As a result, our studies will be suitable for both static and dynamic states.

Now we consider *S(f)* with conditions:

$$\begin{cases} \dfrac{\vec{A_k}^* \vec{L_k}}{m_k} = C \cdots 常数 \\ g_k = g(f) \cdots\cdots k = 1,2,3,\cdots,N \end{cases} \qquad (11)$$

The $\beta_k$ in equation (10) is variable only, then we may replace the sum by integration

$$\sum_k \frac{C}{\sqrt{\beta_k^2 + f^{2\alpha} g(f)}} \propto \int \frac{d\beta}{\sqrt{\beta^2 + f^{2\alpha} g(f)}} \propto \log\left[\sqrt{\beta^2 + f^{2\alpha} g(f)} - \beta\right] \qquad (12)$$

We have

$$S(f) = K \frac{\left| \log\left[ \sqrt{1 + \left(\dfrac{f}{f_0^*}\right)^{2\alpha} g(f)} - 1 \right] \right|}{f} \qquad (13)$$

This is just power spectrum of 1/f noise we want. An achievement of the theoretical analysis is a "characteristic frequency"—constant $f_0^*$ appeared in the equation. It is an important parameter to indicate system character like the tone color of a rigid body.

The power 1/f noise is get from

$$\frac{d}{dt}\int_{-\infty}^{+\infty} S(f)e^{i(\omega-\lambda)t}df = \int_{-\infty}^{+\infty} i(\omega-\lambda)S(f)e^{i(\omega-\lambda)t}df \qquad (14)$$

If $f$ is small, the power spectrum of 1/f noise will be

$$i(\omega-\lambda)S(f) = 2\pi i(f - \frac{\lambda}{2\pi})S(f) \approx -i\lambda S(f) \propto S(f) \qquad (15)$$

Therefore energy spectrum and power spectrum possess same function structure, but the ratio constant $K$ difference.

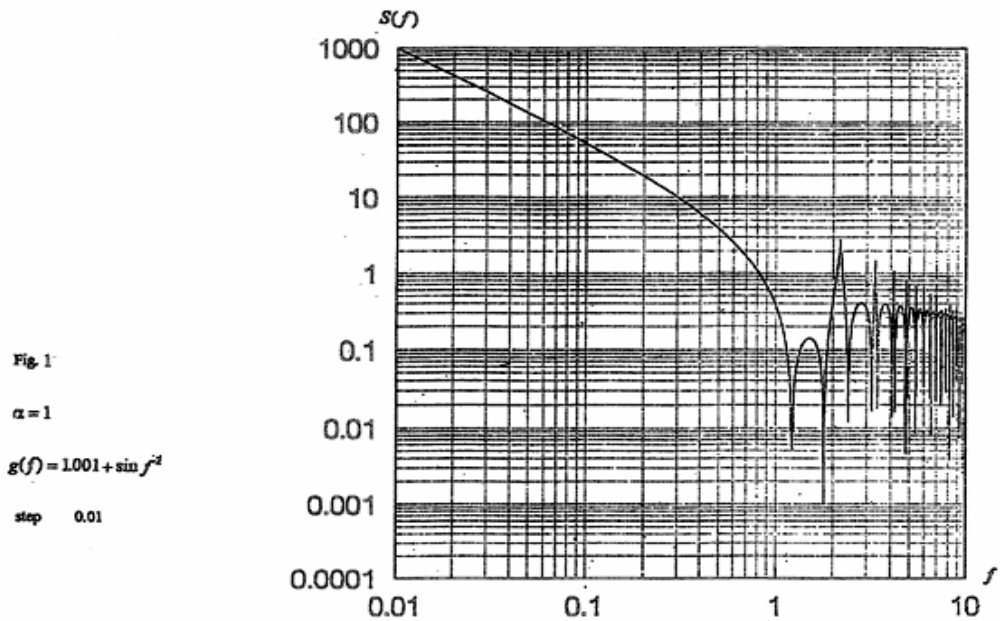

Fig.1 $a=1$, $g(f)=1.01+\sin f^2$, step $0.01$

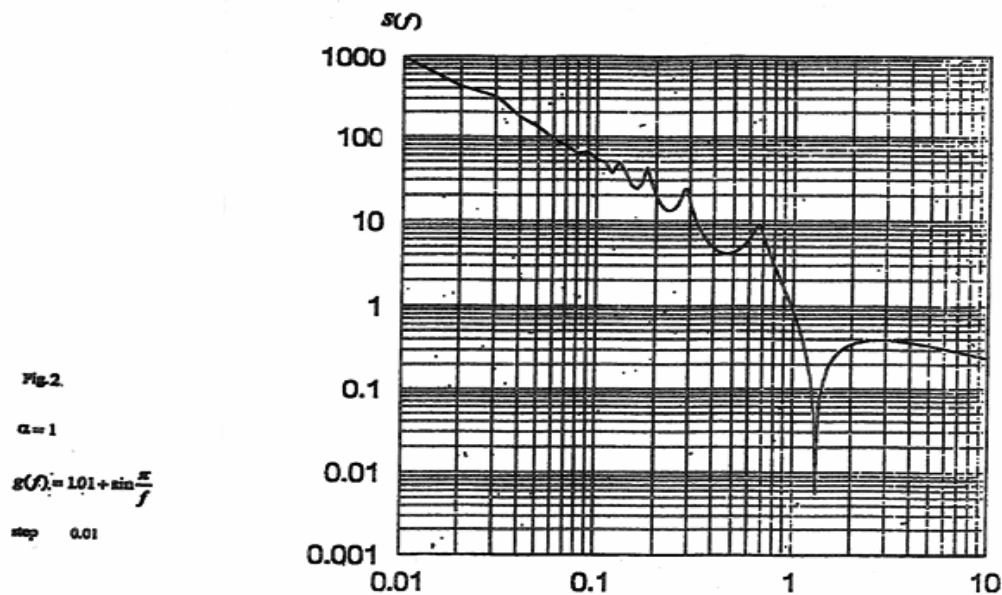

Fig.2 $\alpha = 1$, $g(f) = 1.01 + \sin\frac{\pi}{f}$, step *0.01*

There are three arguments to affirm that equation (13) belongs to 1/f noise. Firstly, equation (13) has included all the known information about 1/f noise; the studies before were concentrated in a small region of $f \to 0$; secondary, equation (13) will solve every problem of 1/f noise; thirdly, as figure 1 and 2 equation (13) has revealed a new and more complex world found never before . Due to we supposed only $\alpha > 1, g(f) > 0$, 1/f noise has infinitive forms.

$K$ in equation (13) is a proportional constant. Experiments shown that the values of $K$ for metals are sensitive to temperature in some regions of temperature[8] while the reference[7] said $K \propto V^\gamma$ for HgCdTe devices, $V$ is operating voltage. Of cause there are still other factors affecting $K$.

According to (13), if
$$x = f/f_0^* \quad (16)$$
much large or small we have
$$S(f) \propto \frac{\left|\log\left(\frac{f}{f_0^*}\right)\right|}{\left(\frac{f}{f_0^*}\right)} \quad (17)$$

Equation (17) is universalism no matter what kind of 1/f noise and dominates all the experiment results.

### 3. The theory compare with the experiment law

Now we would answer the questions as follows:

① Why the power spectra of $\frac{1}{f}$ noise follows the law (1) but not other in various systems such as devices, music, meteorology, pulse of people, crystal wafer oscillation, car flow on the high way *et al*.

② How to explain in theory for difference area as above?

③ The theoretical calculation coincides with experiments well, why?

④ The $\frac{1}{f}$ noise is stable, why?

We begin with a special example. The noise power spectrum for display spectrum measured by Van Vliet etc. has form:
$$S(f) \propto \frac{1}{f^{1.055 \pm 0.015}}$$
Suppose a dimensionless parameter
$$X = \frac{f}{f_0^*}$$
And $\quad Y = \frac{|\log_{10} X|}{X} \quad (18)$

While $2\times 10^{-11} \leq X \leq 5\times 10^{-6}$ we have

$$Z_1 = \frac{|\log_{10} X|}{X} = \frac{6.5832}{X^{1.055}} \qquad (19)$$

The maximum difference of both side of equal-sign is 3.66‰ and standard difference 1.09‰ only. The minimum square imitation for the same data, the result is

$$Z_2 = \frac{6.1716}{X^{1.058677}} \qquad (20)$$

The maximum error between $Z_2$ and $Y$ is 2.23‰, standard error 0.7‰. $Z_2$ approaches $Y$ more than $Z_1$. The values of $Z_1$, $Z_2$, and $Y$ show in table 1, where $\eta = \dfrac{\log_{10} Z - \log_{10} Y}{\log_{10} Y}$.

| X (无量纲频率) | $2\times 10^{-10}$ | 3× | 4× | 5× | 6× | 7× | 8× | $9\times 10^{-10}$ | $1\times 10^{-9}$ | 2 | 3 | 4 | 5 |
|---|---|---|---|---|---|---|---|---|---|---|---|---|---|
| $\log_{10} Y$ (理论值) | 11.0479 | 10.8638 | 10.7332 | 10.6318 | 10.5489 | 10.4787 | 10.4180 | 10.3644 | 10.3164 | 10.0009 | 9.8157 | 9.6843 | 9.5824 |
| $\log_{10} Z_1$ (实验值) | 11.0508 | 10.8650 | 10.7332 | 10.6310 | 10.5474 | 10.4768 | 10.4156 | 10.3617 | 10.3134 | 9.9959 | 9.8100 | 9.6782 | 9.5760 |
| $\eta_1$ (‰) (相对误差) | 0.26 | 0.11 | 0.00 | -0.07 | -0.14 | -0.18 | -0.23 | -0.26 | -0.29 | -0.51 | -0.58 | -0.63 | -0.67 |
| $\log_{10} Z_2$ (计算值) | 11.0585 | 10.8721 | 10.7397 | 10.6372 | 10.5533 | 10.4825 | 10.4211 | 10.3670 | 10.3185 | 9.9998 | 9.8134 | 9.6811 | 9.5785 |
| $\eta_2$ (‰) (相对误差) | 0.96 | 0.76 | 0.61 | 0.51 | 0.42 | 0.36 | 0.30 | 0.25 | 0.20 | -0.11 | -0.23 | -0.33 | -0.41 |

| 6 | 7 | 8 | $9\times 10^{-9}$ | $1\times 10^{-8}$ | 2 | 3 | 4 | 5 | 6 | 7 | 8 | $9\times 10^{-8}$ | $1\times 10^{-7}$ | 2 |
|---|---|---|---|---|---|---|---|---|---|---|---|---|---|---|
| 9.4990 | 9.4285 | 9.3674 | 9.3135 | 9.2653 | 8.9476 | 8.7615 | 8.6293 | 8.5266 | 8.4427 | 8.3717 | 8.3102 | 8.2559 | 8.2073 | 7.8872 |
| 9.4925 | 9.4218 | 9.3606 | 9.3067 | 9.2584 | 8.9408 | 8.7550 | 8.6232 | 8.5210 | 8.4374 | 8.3668 | 8.3056 | 8.2517 | 8.2034 | 7.8852 |
| -0.68 | -0.71 | -0.73 | -0.73 | -0.74 | -0.76 | -0.74 | -0.71 | -0.66 | -0.63 | -0.59 | -0.53 | -0.51 | -0.48 | -0.18 |
| 9.4946 | 9.4238 | 9.3624 | 9.3083 | 9.2598 | 8.9412 | 8.7547 | 8.6224 | 8.5198 | 8.4360 | 8.3651 | 8.3037 | 8.2496 | 8.2011 | 7.8825 |
| -0.46 | -0.50 | -0.53 | -0.56 | -0.59 | -0.72 | -0.78 | -0.80 | -0.80 | -0.79 | -0.79 | -0.78 | -0.76 | -0.76 | -0.60 |

| 3 | 4 | 5 | 6 | 7 | 8 | $9\times 10^{-7}$ | $1\times 10^{-6}$ | 2 | 3 | 4 | $5\times 10^{-6}$ | 均方值 | 标准误差 |
|---|---|---|---|---|---|---|---|---|---|---|---|---|---|
| 7.6995 | 7.5662 | 7.4627 | 7.3780 | 7.3064 | 7.2442 | 7.1894 | 7.1404 | 6.8170 | 6.6273 | 6.4924 | 6.3876 | | |
| 7.7000 | 7.5682 | 7.4660 | 7.3824 | 7.3118 | 7.2506 | 7.1967 | 7.1484 | 6.8308 | 6.6450 | 6.5132 | 6.4110 | | |
| 0.06 | 0.26 | 0.44 | 0.60 | 0.74 | 0.88 | 1.06 | 1.12 | 2.02 | 2.67 | 3.20 | 3.66 | $\eta_1^2 = 1.17867$ | $\sigma_1 = 1.09‰$ |
| 7.6960 | 7.5637 | 7.4611 | 7.3773 | 7.3064 | 7.2450 | 7.1909 | 7.1425 | 6.8238 | 6.6374 | 6.5050 | 6.4024 | | |
| -0.45 | -0.33 | -0.21 | -0.10 | 0.00 | 0.11 | 0.21 | 0.29 | 1.00 | 1.52 | 1.94 | 2.32 | $\eta_2^2 = 0.57792$ | $\sigma_2 = 0.76‰$ |

Table 1 $\quad Y = \dfrac{\log_{10} X}{X}, \quad Z_1 = \dfrac{6.5832}{X^{1.055}} \quad Z_2 = \dfrac{6.1716}{X^{1.058677}} \quad \eta = \dfrac{\log_{10} Z - \log_{10} Y}{\log_{10} Y}$

**Table 1   $Y$, $Z_1$, $Z_2$, and $\eta$**

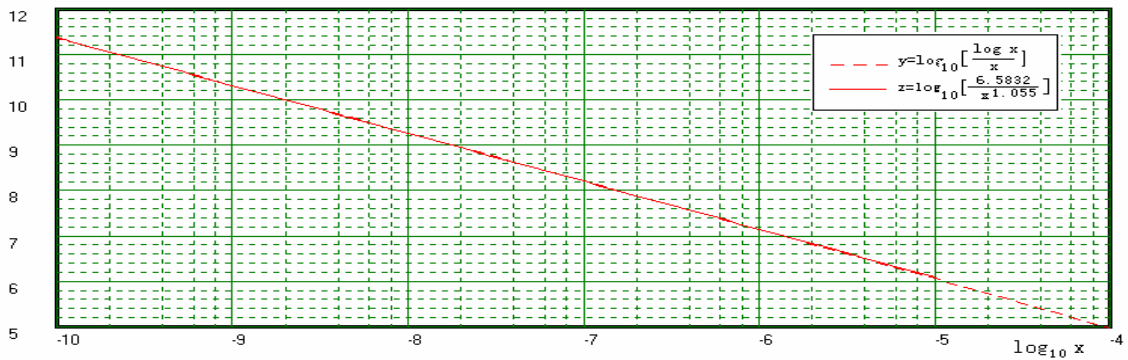

Fig. 3 $\log_{10} Y$ and $\log_{10} Z_1$ vs $\log_{10} X$

$Y$ coincide with $Z_1$ which means that $Z_1$ and $Z_2$ are optimum approximate of $Y$ at least while $2\times 10^{-10} \leqslant X \leqslant 5\times 10^{-6}$.

Now we explain the law (21) for all experimental results of $\frac{1}{f}$ noise.

$$Z = \frac{C}{X^M} \tag{21}$$

In the double log coordinate $Y$ is curve and $Z$ straight line with slope $M$ but almost coincide in especial for certain area. It can be proved. Suppose

$$\xi = -\log_{10} X \quad \text{and} \quad \eta = \log_{10} Y \tag{22}$$

The equation (18) changes to

$$\eta = \xi \pm \log_{10}|\xi| \tag{23}$$

Suppose $\xi = \xi_0 + \Xi = \xi_0\left(1 + \frac{\Xi}{\xi_0}\right) = \xi_0(1+\vartheta)$ \hfill (24)

where $\Xi$ is an increase of $\xi_0$ and as floating coordinate with original point at $\xi_0$, then the expansion of $\eta$ in series at $\xi_0$

$$\eta = \xi_0 + \Xi \pm \log_{10}|\xi_0\|1+\vartheta|$$

$$= \xi_0 \pm \log_{10}|\xi_0| + \left(1\pm\frac{\beta}{\xi_0}\right)\Xi \mp \beta\left(\frac{\vartheta^2}{2} - \frac{\vartheta^3}{3} + \cdots\right) \tag{25}$$

where $\beta = \log_{10} e = 0.4343\cdots$ \hfill (26)

It shows from (25) that the difference $\partial$ of $\eta$ with straight

$$\partial = \beta\left(\frac{\vartheta^2}{2} - \frac{\vartheta^3}{3} + \cdots\right) = \beta[\vartheta - \log_{10}(1+\vartheta)] = \beta\left\{\left(\frac{\Xi}{\xi_0}\right) - \log\left[1+\frac{\Xi}{\xi_0}\right]\right\} \tag{27}$$

The slope of the straight is

$$M = 1 \pm \frac{0.4343}{\xi_0} \tag{28}$$

Therefore the experiment law (1) for all kinds of $\frac{1}{f}$ noise can be explained and the doubts were solved including about $X$, namely $f$ is only continue because the straight line is optimum approximate derivative function of $\eta$. It will see that the value of $M$ got from equation (28) is just one of innumerable permitted values.

We turn to say why $Y$ may be used for numerous fields such as device, music, meteorology *et al*.

The concept "system" means an assembly of units with same characters and possesses boundary maybe not clear which may be space or time. In general, system is a special and explicit concept here. We had separated the matter carrier from system. Therefore the "system" used in theory is extended widely, no matter what is micro or macro, nature or society, including device, music, meteorology pulses of people *et al*. of course. That is to say the equation (13) is suitable for very wide and possesses deep connotation. That is the reason why observed $\frac{1}{f}$ noise follows the law (3). It will be explained with another point of view.

In fact experiment law (1) is derivative function and optimum approximate of $Y$ (equation (18)) known from (25). At same time, $\frac{1}{f}$ noise is stable because there is no time $t$ in (13). It is need to emphasize in special as said before that the value $K$ depends on physical conditions then the maximum of $\frac{1}{f}$ noise should be difference which shows clear or hidden as moonlight in the evening. We will analyze detail in 4.

Usually, one may determine curve as a straight line without (13), then be puzzled by the feature of $\frac{1}{f}$ noise. It is difficulty to reach goal if first step runs wrong, as Bacon said "…"

## 4. The mechanism of $\frac{1}{f}$ noise

Let's to derive the power spectrum (5) of the random force $\vec{\wp(t)}$. Suppose

$$\vec{B}^*(f)\vec{B}(f) = \frac{\vec{L}^*\vec{L}}{G(f)} \tag{29}$$

Make $G(f)$ expansion in Taylor series at $f=o$

$$G(f) = \beta^2 + a_1 f + a_2 f^2 + a_3 f^3 + a_4 f^4 + \cdots = \beta^2 + a_1 f + f^2 g(f) \tag{30}$$

Where $g(f) = a_2 + a_3 f + a_4 f^2 + \cdots$ (31)

The convergence of the integral

$$\int_{-\infty}^{+\infty} \frac{1}{G(f)} df = \int_{-\infty}^{+\infty} \frac{1}{\beta^2 + a_1 f + f^2 g(f)} df \tag{32}$$

needs $a_1=0$, $g(f)>0$, and can be generalized to equation (5). That is to say, there are infinite random act forces. Compared with (13), any random act force with definite α and $g(f)$ will induce a definite kind of $\frac{1}{f}$ noise which is the specialty. Different values of α and $g(f)$ induces different the specialty. So α and $g(f)$ shows generality of $\frac{1}{f}$ noise. It's one characteristic of the theory here that specialty and generality are collected ingeniously in equation (13). Because equation (5) includes all the random act forces, the theory indicates that no matter what random act force must induces $\frac{1}{f}$ noise which therefore can be observed everywhere.

Random act force going in to dynamic equation affects $\frac{1}{f}$ noise. Though each random act force is small, the affection of all the sum of random act force can not be neglected due to there is a lot of random act forces.

The random act forces originate from thermodynamic random fluctuation in system and affects of surroundings out of system. All the things in the world change uninterruptedly and affect the system to form random act force. That is "the surge of background energy" from the view of energy existing random and forever. There is "zero point energy" even at temperature K=0 proved in quantum mechanics

We suppose that the "zero point energy" and "dark matter" existed in the universe may be relative to random act force.

It was reported[15] that I-V character of low frequency noise voltage and noise voltage with bias current measured on Ti film micro-bridge at 77K. The experiments show noise peak usually at critical current and the noise un-changes while cooling sample with difference magnetic field, which means that the fluctuation of critical current is origin of low frequency $\frac{1}{f}$ noise on Ti film micro-bridge tunnel junction and just "the surge of background energy" said as above. The experiment results will be same for Cr, In, ···, not only for Ti[15]. 克拉克 had got the conclusion "the fluctuation of critical current caused by temperature fluctuation" thirty years ago. Except for explaining mechanism of $\frac{1}{f}$ noise, the example shows also important effect of K value (see (13)). The random fluctuation enhances near the critical current and K increases more，then $\frac{1}{f}$ noise becomes obviously, which cant be visible while K small. This mechanism can also be suited to dc SQLUD devices. Other kind of $\frac{1}{f}$ noise is also caused by corresponding fluctuation. Therefore all the particular 'mechanism' may be put into "the surge of background energy", which make knowledge of $\frac{1}{f}$ noise generalize and get great progress.

According to "the surge of background energy", we can explain that the origin of $\frac{1}{f}$ noise is all from interaction between system and random action. Because "the surge of background energy" exists everywhere, $\frac{1}{f}$ noise can be observed in no matter what system belonged to no matter what domain and possesses similar form which coincides with the theory said in paragraph 2. Fortunately "…the fluctuation is related strongly with it's history long ago …" can also be understand. The surge of background energy characterized by special random action must induce certain kind of $\frac{1}{f}$ noise. This means that current state is entangled with past or history portrays today. It is need to explain that "interaction" indicates the random act to go into motion equation and makes system to change. At same time the change of system will affect random act.

Reference [8] indicated that $\frac{1}{f}$ noise can be found in various systems and guessed to originate from a general mechanism, but no satisfied explanation up to now. Our theory here solved the question completely.

The $\frac{1}{f}$ noise can be understood from another point of view.

Supposing a system with energy function $\Phi(t)$ and detecting $\Phi(t)$ with sine wave of frequency $f$, once detection needs time $\tau = \frac{1}{f}$ at least. However $\Phi(t)$ have changed to $\Phi(t+\tau)$, the change of energy function $\Phi(t)$

$$\Delta\Phi = \Phi(t+\tau) - \Phi(t) = \frac{\partial \Phi}{\partial t}\tau \tag{33}$$

If a random fluctuation around the average $\Delta\Phi$ is just noise and the power spectrum is

$$S(f) = \frac{\partial \Delta\Phi}{\partial t} = \frac{\partial^2 \Phi}{\partial t^2}\tau \tag{34}$$

For multi-detections

$$S(f)_{t_k} = \left(\frac{\partial^2 \Phi}{\partial t^2}\right)_{t_k} \tau \tag{35}$$

$t_k$ is begin time of $k$th detection and $k=1,2,3, \cdots N$

The average square root of equation (35)

$$S(f) = \sqrt{\frac{1}{N}\sum_{k=1}^{N} S^2(f)_{t_k}} = B\tau = \frac{B}{f} \tag{36}$$

where $B = \sqrt{\frac{1}{N}\sum_{k=1}^{N}\left[\left(\frac{\partial^2 \Phi}{\partial t^2}\right)_{t_k}\right]^2}$ (37)

It is clear from (36) that the $f$ less the detection time $\tau$ then $S(f)$ longer which is a special state for $M=1$. The $f$ contradicts $S(f)$, therefore $\frac{1}{f}$ noise can be as a "uncertainty relation" in macroscopic limitedly.

## 5. Analyzing M value

We have explained that the experiment law of $\frac{1}{f}$ noise must follow equation (1) and the mechanism of $\frac{1}{f}$ noise. Now we turn to analyze the value of M being not 1/2,1,2,4,⋯, as in general physical equation and regular seemingly but not in fact, which has puzzled people for 80 years. For solving the question we begin with physical area $1.0\times10^{-3}\leqslant x\leqslant 1.0\times10^{-2}$ as in reference [14] and sample (measure) at $1.0\times10^{-3}$, $1.5\times10^{-3}$, $2.0\times10^{-3}$, ⋯⋯, $9.5\times10^{-3}$, $1.0\times10^{-2}$, total 19 sampling points. The values $Z_1$, $Z_2$, $Z_3$, $Z_4$, $Z_5$, list in table2.

|  | Standard error | Sampling number |  |
|---|---|---|---|
| $Z_1 = \frac{0.88609}{x^{1.17895}}$ | $\sigma_1=0.00290$ | n=19 | One $Z_1$ only |
| $Z_2 = \frac{0.88921}{x^{1.17821}}$ | $\sigma_2=0.00300$ | n=17 | 171 different $Z_2$ |
| $Z_3 = \frac{0.88971}{x^{1.17811}}$ | $\sigma_3=0.00310$ | n=15 | 3876 $Z_3$ |
| $Z_4 = \frac{0.88815}{x^{1.17843}}$ | $\sigma_4=0.00313$ | n=13 | 27132 $Z_4$ |
| $Z_5 = \frac{0.89246}{x^{1.17746}}$ | $\sigma_5=0.00324$ | n=10 | 92378 $Z_5$ |
| total |  |  | 123558 |

Table 2. The values $Z_1$, $Z_2$, $Z_3$, $Z_4$, $Z_5$

| $x$ | | $1.0\times10^{-3}$ | $1.5\times10^{-3}$ | $2.0\times10^{-3}$ | $2.5\times10^{-3}$ | $3.0\times10^{-3}$ | $3.5\times10^{-3}$ | $4.0\times10^{-3}$ | $4.5\times10^{-3}$ | $5.0\times10^{-3}$ | $5.5\times10^{-3}$ |
|---|---|---|---|---|---|---|---|---|---|---|---|
| $\log_{10}Y$ (理论值) | | 3.47712 | 3.27476 | 3.13017 | 3.01738 | 2.92478 | 2.84615 | 2.77778 | 2.71726 | 2.66295 | 2.61368 |
| 计算值 | $\log_{10}Z_1$ | 3.48433 | 3.27673 | 3.12943 | 3.01518 | 2.92183 | 2.84290 | 2.77453 | 2.71423 | 2.66028 | 2.61148 |
| | $\partial_1$ | 0.00721 | 0.00197 | −0.00074 | −0.00220 | −0.00295 | −0.00325 | −0.00325 | −0.00303 | −0.00267 | −0.00220 |
| | $\log_{10}Z_2$ | 3.48381 | 3.27633 | 3.12912 |  | 2.92163 | 2.84275 | 2.77442 | 2.71415 | 2.66023 | 2.61147 |
| | $\partial_2$ | 0.00669 | 0.00157 | −0.00105 |  | −0.00315 | −0.00340 | −0.00336 | −0.00311 | −0.00272 | −0.00221 |
| | $\log_{10}Z_3$ | 3.48358 | 3.27613 |  | 3.01476 | 2.92148 | 2.84261 | 2.77429 |  | 2.66012 | 2.61135 |
| | $\partial_3$ | 0.00646 | 0.00137 |  | −0.00262 | −0.00330 | −0.00354 | −0.00349 |  | −0.00283 | −0.00233 |
| | $\log_{10}Z_4$ | 3.48378 |  | 3.12904 | 3.01484 |  | 2.84263 | 2.77429 |  | 2.66009 | 2.61133 |
| | $\partial_4$ | 0.00666 |  | −0.00113 | −0.00254 |  | −0.00352 | −0.00349 |  | −0.00286 | −0.00235 |
| | $\log_{10}Z_5$ | 3.48297 |  | 3.12852 |  | 2.92118 |  | 2.77407 |  | 2.65996 |  |
| | $\partial_5$ | 0.00585 |  | −0.00165 |  | −0.00360 |  | −0.00371 |  | −0.00299 |  |

| 6.0×10⁻³ | 6.5×10⁻³ | 7.0×10⁻³ | 7.5×10⁻³ | 8.0×10⁻³ | 8.5×10⁻³ | 9.0×10⁻³ | 9.5×10⁻³ | 1.0×10⁻³ | 标准误差 | M | C |
|---|---|---|---|---|---|---|---|---|---|---|---|
| 2.56857 | 2.52696 | 2.48833 | 2.45229 | 2.41849 | 2.38667 | 2.35662 | 2.32812 | 2.30103 | | | |
| 2.56693 | 2.52595 | 2.48800 | 2.45268 | 2.41963 | 2.38859 | 2.35933 | 2.33165 | 2.30538 | 0.00290 | 1.17895 | 0.88609 |
| −0.00164 | −0.00101 | −0.00033 | 0.00039 | 0.00114 | 0.00192 | 0.00271 | 0.00353 | 0.00435 | | | |
| 2.56694 | 2.52598 | 2.48805 | 2.45275 | | 2.38870 | 2.35946 | 2.33179 | 2.30554 | 0.00300 | 1.17827 | 0.88921 |
| −0.00163 | −0.00098 | −0.00028 | 0.00046 | | 0.00203 | 0.00284 | 0.00367 | 0.00451 | | | |
| 2.56683 | | 2.48796 | 2.45266 | 2.41964 | | 2.35938 | 2.33172 | 2.30547 | 0.00310 | 1.17811 | 0.88971 |
| −0.00174 | | −0.00037 | 0.00037 | 0.00115 | | 0.00276 | 0.00360 | 0.00444 | | | |
| | 2.52582 | 2.48787 | | 2.41955 | 2.38852 | | 2.33159 | 2.30535 | 0.00313 | 1.17843 | 0.88815 |
| | −0.00114 | −0.00046 | | 0.00106 | 0.00185 | | 0.00347 | 0.00432 | | | |
| 2.56673 | | 2.48790 | | 2.41962 | | 2.35939 | | 2.30551 | 0.00324 | 1.17746 | 0.89246 |
| −0.00184 | | −0.00043 | | 0.00113 | | 0.00277 | | 0.00448 | | | |

Table 3. The values of Y, $Z_1, Z_2, Z_3, Z_4$ and $Z_5$. $\partial = \log_{10} Z - \log_{10} Y$

The values of $Z_1$, $Z_2$, $Z_3$, $Z_4$ and $Z_5$ in table 3 were got through selecting 19,17,15,13 and 10 data from 19 sampling points and by method of least squares. All the standard errors are near 0.00300 and it is difficulty to determine which one of Z's is in accord with theory. The values of M located in a small area are so much to distinguish which is more correct and the difference between two neighbor values of M may be less than any small value.

Before knowing Y existed people considered that the value of M was independent and selected different area e.g. 1.17746<M<1.17843 or 1.17811<M<1.17895 etc. In fact they are different approximate value of Y in same area. Therefore we are going to the results about characters of M as follows.

① The values of M located in a small area are so many to distinguish which one is more correct，no one is special. Therefore M is said to be 'constant' but not stable solid value and must locate in an open area.

② The value of M can not select 1/2,1,2,4, ⋯ otherwise there is large error between function $\frac{C}{f^M}$ and experiment no matter what $C$ is. That is to say an un-proper line replaces Y in double log. Coordinate.

③ We can not determine which one of M values is located at edge so the area in which M existed must be open.

④ M can not be in un-continues area, e.g. $M_1' < M < M_1''$, $M_2' < M < M_2''$, and $M_1'' < M_2'$. We will prove that the area M located must continues.

⑤ Among so many values of M which one is reliable? The answer is that all of them are reliable due to $Z_1, Z_2, Z_3, \cdots\cdots$ are derivative functions, but no one complete because all of them are different from Y with any small value. M changes continuously. Y almost is a straight line special in a certain area $x_1 < x < x_2$ showing in fig.4 and both straight line AA and BB are approximations to Y.

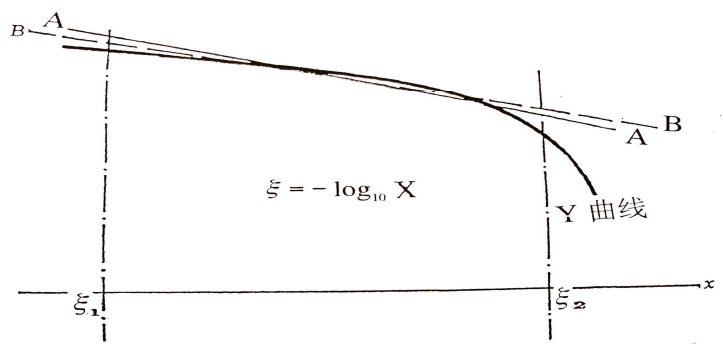

Fig. 4 Curve Y approximated with straight line

There are infinite such lines with different slopes-M which can not be limited to $1/2, 1, 2, 4, \cdots$ of course. In fact M can select any value in an area. From mathematical analysis the straight line AA may change to BB continuously and vice versa. Therefore it has been proved that M changes continuously including all separate values said as above.

⑥ People can not distinguish which value is "edge" among various values and determine the ends of the open area. Therefore many groups may obtain different results. We had given a sample before.

We had solved the question about strange characters of M. The value M represents the quintessence of the question but accuracy is not enough.

## 6. The other things need to discuss

Why M is different for two samples with the same matter and structure.

We discuss crystal as an example. In fact there is no complete perfect crystal without any defects. The character frequency $f_0^*$ must be different for two devices made with same matter, suppose as $f_1^*$ and $f_2^*$. From (22)

$$\Im = \log Z = L - M \log\left(\frac{f}{f_0^*}\right) \tag{38}$$

then

$$X_s = \frac{f_s}{f_1^*} \quad X_s^* = \frac{f_s}{f_2^*} \quad s = 1, 2, \cdots\cdots, n \tag{39}$$

are different due to different $f_1^*$ and $f_2^*$ even $\Im_1, \Im_2, \Im_3, \cdots$ measured at frequency $f_1, f_2, f_3, \cdots$ are all the same. A group of $\Im$ may be corresponding to two groups $X_s$ and $X_s^*$ which is possible? while *L* and *M* select different values only as said in (19) and (20). The specific characteristics cant be ignored in 'particular state' for fascinating $f_0^*$. In fact $f_0^*$, α and *g(f)* play supporting roles only in theory.

The sum was replaced by infinite integral in paragraph 2 and the items of sum were not limited. The results are same no matter how many items with vague logic which represents $\frac{1}{f}$ noise is a general phenomenon. It is the key and difficulty to extract power spectrum from infinite integral which needs knowledge and skill.

General speaking the value of M shown in equation (2) measured by different groups is different and may be explained by equation (29). Then the question about M was solved. The $\frac{1}{f}$ noise follows exact law and its power spectrum and energy spectrum possesses clear and similar structure equations uniquely.

Now we focus our attention on equation (13). The fact $\alpha \geq 1$ and $g(f) > 0$ in equation (13) can reveal many kinds of $\frac{1}{f}$ noise un-found before. Therefore new theory has adaptability widely. It is need to emphasize that the experiment law (1) can be inducted from (17) easily but it is not the same the other way round that the equation (13) can't be inducted from (1).

We turn to the constant K affecting peak of $\frac{1}{f}$ noise in (13). There are special methods to reduce K then $\frac{1}{f}$ noise for various areas which have important interests.

While *f* or $f_0^*$ is vary small then

$$s(f) \propto \frac{|\log f|}{f} \qquad (40)$$

The equation (40) without $f_0^*$, *g(f)*, and α is a universal relationship no matter what form of $\frac{1}{f}$ noise and what system including surroundings, and indicates that the $\frac{1}{f}$ noise exists everywhere and the forms of power spectrum are like. Therefore the generality of $\frac{1}{f}$ noise is very clear.

The equation (40) is as simple and beautiful as Newton's law $\vec{F} = m\vec{a}$,

universal gravitational law $F = K\frac{mM}{R^2}$, Fermat principle $\delta\int_a^b nds = 0$, Planck quantum equation $E = h\nu$, Einstein matter-energy equation $E = mc^2$, Heisenberg uncertainty relation $\Delta P \cdot \Delta r \geq \frac{1}{2}\hbar$ and de Broglie matter wave $\lambda = \frac{h}{P}$ and so on.

In attention to the conclusions as above some suggestions need to discussion:

① There are two conditions used in paragraph 2 what is their connotation?
② If $s(f)$ changes in measurement at different temperature $T$ e.g. room, liquid nitrogen and liquid helium, $s(f)$ should be $s(f,T)$.
③ It is concerned whether "zero energy" and "dark matter" will affect "the surge of background energy".

E-mails: xsj19710726@sina.com or xsj19710726@163.com